\documentclass[pra,aps,preprint]{revtex4-1}
\usepackage{amsmath}
\usepackage{graphicx,mathrsfs}
\usepackage{epstopdf}
\usepackage{epsfig}
\usepackage{xcolor}

\begin{document}

\title{Fine structure of the exciton absorption in semiconductor
superlattices in crossed electric and magnetic fields}

\author{B.~S.~Monozon}
\affiliation{Physics Department, Marine Technical University,
3~Lotsmanskaya Str.,
190121 St.-Petersburg, Russia}

\author{V.~G.~Bezchastnov}
\altaffiliation[On leave from: ]{Ioffe Physical-Technical Institute,
194021 St.-Petersburg, Russia}
\altaffiliation[Present address: ]{Max Planck Institute for Medical Research,
Jahnstrasse 29, 69120 Heidelberg}
\affiliation{Institute of Physical Chemistry, University of Heidelberg, INF 229,
             69120 Heidelberg, Germany}

\author{P.~Schmelcher}
\affiliation{Zentrum f\"ur Optische Quantentechnologien, Universit\"{a}t Hamburg,
Luruper Chaussee 149, 22761 Hamburg, Germany}
\affiliation{The Hamburg Centre for Ultrafast Imaging, Universit\"{a}t Hamburg,
Luruper Chaussee 149, 22761 Hamburg, Germany}

\date{\today}

\begin{abstract}
The
exciton absorption coefficient is determined analytically for a
semiconductor superlattice in crossed electric and magnetic fields,
for the magnetic field being parallel and the electric field being
perpendicular to the superlattice axis.
Our investigation applies to the case where the magnetic length,
while being much smaller
than the exiton Bohr radius, considerably exceeds the superlattice period.
The optical absorption in superlattices displays a spectral fine structure
related to the sequences of exciton states bound whose energies are
adjacent to the Landau energies of the charge carriers
in the magnetic field.
We study effects of external fields and of the centre-of-mass
exciton motion on the
fine structure peak positions and oscillator strengths.
In particular, we find that
the inversion of the orientation of the external fields
and of the in-plane total exciton momentum notably affects the absorption spectrum.
Conditions for the experimental observation of the exciton absorption are discussed.

\end{abstract}

\maketitle

\section{Introduction}\label{S:intro}

Semiconductor Superlattices (SL) are since a few decades
in the focus of intense theoretical and experimental research.
Starting from the studies by Esaki and Tsu \cite{esaki}, a vivid interest
exists in the electronic, transport and optical properties of SLs.
The presence of external electric $\vec{F}$ and magnetic $\vec{B}$ fields
is known to strongly influence these properties, in particular
due to the electric Wannier-Stark (WS) \cite{wann}
and the magnetic Landau \cite{land} quantizations of the electronic motion.

Different field orientations have been considered for the situations where SL
are exposed to either an electric or a magnetic field, or to both fields.
For example, the quantization effects in the optical response of SL were investigated
for $\vec{F} \parallel \vec{e}_z$ in Refs.~\cite{bleus,dig,mend,agul} and for
$\vec{B} \parallel \vec{e}_z$ in Refs.~\cite{maan,maan1},
where the unit vector $\vec{e}_z$ determines the direction of the SL axis.
Combined effects of both fields on the interband optical transitions were
studied by Pacheco \emph{et.al.}~\cite{pach} for
$\vec{F} \parallel \vec{B} \parallel \vec{e}_z$. Suris \emph{et.al.} calculated
the energy spectrum \cite{ber}, optical absorption coefficient \cite{ber1} and
magnetoresistivity \cite{shch} for crossed fields - longitudinal electric,
$\vec{F} \parallel \vec{e}_z$, and in-plane magnetic field $\vec{B} \perp \vec{e}_z$.
For the same field geometry, Zuleta and Reyes-Gomes~\cite{zurey} have recently
investigated the interband optical absorption spectra of a GaAs/GaAlAs variably
spaced SL.

According to the experimental measurements~\cite{nel,chom},
the electronic and optical properties of quasi-2D semiconductor
structures are notably influenced by the excitons - the
electron-hole pairs bound by the Coulomb attraction.
For SLs exposed to parallel ($\vec{B} \parallel \vec{F} \parallel \vec{e}_z$)
and crossed ($\vec{F} \parallel \vec{e}_z$, $\vec{B} \perp \vec{e}_z$) fields,
exciton effects were considered in Ref.~\cite{pach1} and
Ref.~\cite{dig1}, respectively. A variety of studies
on SL excitons in external fields are described in the
reviews and monographs~\cite{bast,barn,glut}.

Despite
the fact that
basic properties of SL excitons are comprehensively
studied, some qualitatively novel results have been reported recently.
Suris~\cite{suris} has revealed and
analytically investigated
an interplay between the center-of-mass and relative motions
of exitons in a semiconductor SL. This effect
results from the deviation of the
dispersion for the exciton electron-hole pair from
a quadratic law
and leads to a fine structure of the
exciton Rybderg states.
A similar structure was studied earlier by Kohn and Luttinger \cite{kohnlutt} for
the Ge and Si bulk crystals with extremely anisotropic isoenergy band surfaces.

The works \cite{suris,kohnlutt}
do not address situations where external fields are applied to SL. For a
longitudinal electric field, $\vec{F} \parallel \vec{e}_z$,
the fine structure of the exciton absorption was studied
in Ref. \cite{monschm17}.
The binding energies $E^{({\rm b})}_n$ and the oscillator strengths $f_n$
of the fine structure states $n=0,1,2,\ldots$
were shown to be of the order of $4{\rm Ry}/(2n + 1)^2$ and $1/(a_0^2\bar{z})$,
respectively, where ${\rm Ry}\sim a_0^{-2}$ is the exciton Rydberg constant,
$a_0$ is the exciton Bohr radius, and
$\bar{z} \sim d^{\,2/3}(\Delta_e + \Delta_h)^{1/3}$ is the longitudinal
size of the exciton. This size is determined by
the SL period $d$ and the miniband widths $\Delta_e$ for the electrons and
$\Delta_h$ for the holes, with the widths being dependent on the longitudinal
exciton momentum in the electric field applied.

The studies \cite{suris,monschm17} employ
the adiabatic, $2\bar{z} \ll a_0$, and
the continual, $d \ll \bar{z}$, approximations which facilitate the analytical
calculations. Justifying both approximations requires
the condition $2d \ll a_0$
which, as already pointed out in Ref.~\cite{suris}, might not be well satisfied
in typical SLs. For example, for the GaAs/AlGaAs SL with a period of $2$~nm, the
Bohr radius $a_0=11.4$~nm does not significantly exceed the value $2d=4$~nm
to provide the adiabatic and continual approximations.
On the other hand, with increasing $a_0$ the exciton binding energies and
oscillator strengths decrease being both proportional to $a_0^{-2}$. This
makes the crystals with larger values for the exciton Bohr radius less
suitable for an experimental probing of the fine structure. However,
the presence
of an external magnetic field can increase both the binding energies and the oscillator
strengths making thereby the fine spectral structure more pronounced as
compared to the case without
the field. Theoretical calculations of the fine structure in the
presence of the external fields are therefore of fundamental interest and
relevant for supporting the experimental studies.

In the present paper we consider a SL of a short period $d \ll a_0$ subject
to crossed external electric $\vec{F}$ and magnetic $\vec{B}$ fields.
We focus on an analytical investigation
and analysis of the exciton bound states and optical absorption.
The external magnetic field is assumed to be stronger then the  exciton internal
field providing the magnetic length $a_B = \sqrt{\hbar/(eB)}$ to be much
smaller than the exciton Bohr radius $a_0$. Thus, for the magnetic field directed
along the SL axis, we study a regime where the in-plane exiton size $\sim a_B$
becomes much smaller than this size $\sim a_0$ in the absence of the field
and leads to a significant increase of the fine structure binding energies and
oscillator strengths. We show that the presence on an in-plane electric field
results in additional exciton peaks in the magneto-absorption spectrum.
The impact of the combined magnetic and electric fields on
the exciton bound states is studied in detail. In particular, we investigate a
coupling between the longitudinal and in-plane collective motions of the exciton,
the internal motion, and external fields. To deliver close-form analytical
results we consider a simple
band model of SL and employ tight-binding, adiabatic
($\bar{z}\ll a_B \ll a_0$) and continual ($d\ll \bar{z}$) approximations
for the exciton states and corresponding optical absorption coefficients.

The paper is organized as follows.
The theoretical framework to describe
the exciton states for the SL in the presence of
crossed electric and magnetic fields is given in Section 2.
The energies and wave functions of the bound states are presented in
explicit form in Section 3.
In Section 4 we determine the absorption coefficient and analyze its dependencies
on the SL parameters, on the external field strengths, and on the total exciton momentum.
The applicability of the obtained results is discussed and expected experimental values
are estimated. Our concluding remarks are given in Section 5.

\section{General approach}\label{S:gen}

The Hamiltonian describing a Wannier-Mott exciton has the form
\begin{equation}
H_{\rm ex} = \sum_j H_j + U,
\label{H_ex}
\end{equation}
where $j$ designates an electron ($j=e$) and a hole ($j=h$) with the
effective masses $m_j$ and the charges $e_j$ ($e_e = -e_h = -e$)
linked to each other by an attractive Coulomb potential
\begin{equation}
U = -\frac{1}{4\pi\epsilon_0\epsilon}\cdot\frac{e^2}{r},
\label{U}
\end{equation}
with $\epsilon$ being a semiconductor dielectric constant, and
$r=|\vec{r}_e-\vec{r}_h|$ being the distance between the electron
and hole located at
$\vec{r}_e$ and $\vec{r}_h$, respectively. The Hamiltonian~(\ref{H_ex})
and the effective mass approximation apply to semiconductors with
parabolic, non-degenerate, and spherically symmetric energy bands separated by
a wide energy gap $E_g$.

In a semiconductor SL exposed to uniform external fields
the Hamiltonians $H_j$ are
\begin{equation}
H_j = -\frac{\hbar^2}{2m_j}\frac{\partial^2}{\partial z_j^2}
      + V_j + H_{\perp j} + e_j\vec{F}\vec{\rho}_j,
\label{H_j}
\end{equation}
where $z_j$ and $\vec{\rho}_j$ are the longitudinal and transverse components
of $\vec{r}_j = (\vec{\rho}_j,z)$ with respect to the SL axis,
$V_j = V_j(z_j)$ are the potentials of the SL layers, and
\begin{equation}
H_{\perp j} = \frac{1}{2m_j}
              \left( -{\rm i}\hbar \frac{\partial}{\partial \vec{\rho}_j}
                     + \frac{e_j}{2} \vec{B} \times \vec{\rho}_j \right)^2
\label{H_tr_j}
\end{equation}
are the kinetic energies of the motions of the electron and the hole transversely
to the magnetic field.
The above equation explicitly accounts for the field orientations
$\vec{F} \perp \vec{e}_z$ and $\vec{B} \parallel \vec{e}_z$, and the
symmetrical gauge of the magnetic vector-potential is employed in
$H_{\perp j}$.

We follow an approach described in details in Refs.~\cite{suris,monschm17} and
present below a brief outline of intermediate calculations.
The SL potentials are periodic functions
$V_j(z_j)= V_j(z_j + n_j d)$ with the period $d$
formed by a large number $n_j = 0,1,\ldots N_0$
of quantum wells separated by weakly penetrable barriers.
We assume the size-quantized energies $b_j \sim \hbar^2/(m_j d^2)$
to significantly exceed the miniband widths $\Delta_j$ and apply a nearest-neighbor
tight-binding approximation for the inter-well tunneling.

Expanding an eigenfunction of the Hamiltonian~(\ref{H_ex}) in a set of
orthonormalized Wannier functions $w_j(z_j - n_j d)$,
\begin{equation}\label{E:expan}
\Psi (\vec{r}_e, \vec{r}_h) = \sum_{n_e,n_h}
\Phi(n_e,\vec{\rho}_e;n_h,\vec{\rho}_h)w_e(z_e-dn_e)w_h(z_h-dn_h),
\end{equation}
and exploiting conservation of the exciton transverse $\vec{K}$ and longitudinal
$Q$ momenta, one can obtain
\begin{equation}\label{E:Phi}
\Phi(n_e,\vec{\rho}_e;n_h,\vec{\rho}_h)
=e^{{\rm i}\alpha\left(\vec{R}_{\perp},\vec{\rho}\right)}\,g(n,\vec{\rho}),
\end{equation}
where $\vec{R}_{\perp} = (m_e \vec{\rho}_e + m_h \vec{\rho}_h)/M$ and
$\vec{\rho} = \vec{\rho}_e - \vec{\rho}_h$ are the center-of-mass and relative
transverse coordinates, respectively, $M = m_e + m_h$ is the exciton mass,
$n=n_e-n_h$,
\begin{align}\label{E:alpha}
&
\alpha(\vec{R}_\perp,\vec{\rho}) =
\left(\vec{K}+\frac{e}{2\hbar}\vec{B}\times\vec{\rho}\right)\!\vec{R}_\perp
+ \frac{\delta}{2}\,\vec{K}'\vec{\rho}
+ \frac{Qd}{2}\,(n_e+n_h)
- {\gamma}dn,
\nonumber \\
&
\delta = \frac{m_h - m_e}{M}, \;\;\;
\gamma = \frac{1}{d}\,
             \tan^{-1}\left[\frac{\Delta_e -\Delta_h}{\Delta_e + \Delta_h}
                            \tan \left(\frac{Qd}{2}\right)\right],
\nonumber \\
&
\vec{K}' = \vec{K} + \frac{M}{\hbar B^2}\,\vec{B}\times\vec{F}.
\end{align}

We consider a regime $a_B \ll a_0$ where the external magnetic field dominantly
determinesthe internal exciton motion in the coordinate $\vec{\rho}$
and the function $g(n,\vec{\rho})$ has the form
\begin{equation}\label{E:g}
g(n,\vec{\rho}) = \Lambda_{Nm}(\vec{\rho}-\vec{\rho}_0)\,\psi(n),
\end{equation}
where $\Lambda_{Nm}(\vec{\rho}-\vec{\rho}_0)$ with $N = 0,1,2,\ldots$
and $m = 0,\pm 1,\pm 2,\ldots$ are
the Landau functions describing relative motion of the charges $\pm e$ in the
magnetic field with respect to the center
\begin{equation}
\vec{\rho}_0 = \frac{\hbar}{eB^2}\,\vec{B}\times\vec{K}'.
\label{rho_0}
\end{equation}
The set $\Lambda_{Nm}$ is orthonormalized,
$\langle \Lambda_{Nm} \mid \Lambda_{N'm'} \rangle = \delta_{NN'}\,\delta_{mm'}$,
and is given by the expressions
\begin{eqnarray}\label{E:radial}
\Lambda_{Nm}(r_\perp, \varphi) &=&
\frac{{\rm e}^{{\rm i} m\varphi}}{\sqrt{2\pi}}\,R_{Nm}(r_\perp),
\nonumber \\
R_{Nm}(r_\perp) &=& \frac{1}{a_B}\,\sqrt{\frac{N!}{(N+\mid m \mid)!}}\,
u^{\textstyle \frac{\mid m \mid}{2}}\,
{\rm e}^{\textstyle -\frac{u }{2}}\,
L_N^{\mid m \mid}(u),
\end{eqnarray}
where
$r_\perp^2 = \left(\rho_x-\rho_{0x}\right)^2 + \left(\rho_y-\rho_{0y}\right)^2$,
$\varphi=\tan^{-1}\left[\left(\rho_y-\rho_{0y}\right)/
                        \left(\rho_x-\rho_{0x}\right)\right]$,
$u=r_\perp^2/(2a_B^2)$, and $L_N^{\mid m \mid}$ are the associated Laguerre
polynomials.

The search for an eigenfunction of the Hamiltonian~(\ref{H_ex}) in the form determined
by Eqs.~(\ref{E:expan}), (\ref{E:Phi}) and (\ref{E:g}) leads to the equation
for the function $\psi(n)$ and energies $W_{Nm}$:
\begin{equation}\label{E:relat}
\frac{\Delta_{eh}}{4} \Big[ 2\psi(n) - \psi(n+1) - \psi(n-1) \Big] +
\Big[ \bar{U}_{Nm}(n;\rho_0) - W_{Nm} \Big] \psi(n) = 0,
\end{equation}
where $\Delta_{eh}$ is a reduced electron-hole miniband given by the
relation~\cite{suris},
\begin{equation}\label{E:mini}
\Delta_{eh}^2 = (\Delta_e + \Delta_h)^2 -2 \Delta_e \Delta_h \left[ 1 - \cos(Qd) \right],
\end{equation}
$\bar{U}_{Nm}(n;\rho_0)$
is the average of the electron-hole interaction potential~(\ref{U}) with
the product of the densities
$ \left| w_e(z_e-dn_e) \right|^2 $,
$ \left| w_h(z_h-dn_h) \right|^2 $, and
$ \left| \Lambda_{N,m}(\vec{\rho}-\vec{\rho}_0) \right|^2 $,
and $W_{Nm}$ determines the exciton energy
\begin{equation}\label{E:energy}
E = \mathscr{E}_g + T(Q)
  + \frac{\hbar^2}{2M}\,\left( K^2 - {K'}^2 \right)
  + \frac{\hbar eB}{2\mu}\,\Big( 2N + |m| + \,\delta\cdot m + 1 \Big)
  + W_{Nm}.
\end{equation}
In Eq.~(\ref{E:energy}), $\mathscr{E}_g = E_g + b_e + b_h$ is the SL energy gap
($b_e$ and $b_h$ are the size-quantized energies of the electron and the hole,
respectively),
\begin{equation}\label{E:TQ}
T(Q)= \frac{1}{2} \Big[ \Delta_e + \Delta_h - \Delta_{eh}(Q) \Big]
\end{equation}
is the energy of the center-of-mass longitudinal motion, whereas the third and fourth
terms describe the energies of Landau $Nm$-levels shifted due to the effects of
the in-plane electric field $\vec{F}$ and the exciton motion with the momentum $\vec{K}$.

For the exciton states with the longitudinal size exceeding significantly the SL
period, $\bar{z} \gg d$, one can replace
\begin{equation}\label{E:replacements}
nd \to z, \;\;\;
\psi(n+1) + \psi(n-1) -2\psi(n) \to d^2\,\psi''(z),
\end{equation}
where $z = z_e - z_h$ is the exciton relative coordinate.
This establishes a continual approximation to Eq.~(\ref{E:relat}):
\begin{equation}\label{E:relat1}
\psi''(z) +
\frac{2m_{\parallel}(Q)}{\hbar^2}
\Big[ W_{Nm} - \bar{U}_{Nm}(z;\rho_0) \Big] \psi(z) = 0,
\end{equation}
where
\begin{equation}\label{E:m_long}
m_{\parallel}(Q) = \frac{2\hbar^2}{d^2\Delta_{eh}(Q)}
\end{equation}
has the meaning of a reduced ``longitudinal'' effective mass of the exciton determined
by its longitudinal momentum $Q$, and
\begin{equation}\label{E:pot1}
\bar{U}_{Nm}(z;\rho_0)=
\frac{2}{\pi}\int_0^{\infty}\rho d\rho \, R_{Nm}^2 (\rho) \,
\frac{F(\pi/4,k)+F(\phi,k)}{\sqrt{(\rho + \rho_0)^2 +z^2}}
\end{equation}
is an effective binding potential.
The function
$F(\phi,k)$ is an elliptic integral of first kind \cite{ryzgr},
\begin{equation}\label{E:pot2}
\phi = \sin^{-1}\sqrt{\frac{(\rho +\rho_0)^2 +z^2}{2(\rho^2 +\rho_0^2 +z^2 )}},
\;\;\;
k^2 = \frac{4\rho \rho_0}{(\rho +\rho_0)^2 +z^2},
\;\;\;
\rho_0 = \sqrt{\rho_{0x}^2+\rho_{0y}^2}.
\end{equation}

In the following we study bound solutions of Eq.~(\ref{E:relat1})
distinguished by a discrete quantum number $p=0,1,2,...$.
Having obtained the bound-state energies $W_{Nm} = W_{Nmp}$ and the
corresponding functions $\psi_p(z)$,
$\langle \psi_p \mid \psi_{p'} \rangle = \delta_{pp'}$, we
can determine the exciton total energy
$E = E_{\vec{K}QNmp}$ according to Eq.~(\ref{E:energy}) and
determine the wave functions
\begin{align}\label{E:totfun}
\Psi_{\vec{K}QNmp}\left(\vec{r}_e,\vec{r}_h\right) &=
d \sum_{n_e,n_h} w(z_e - dn_e) w(z_h - dn_h)\,
\frac{e^{{\rm i}\beta_{\vec{K}Q}\left(\vec{r}_e,\vec{r}_h\right)}}{\sqrt{SL}}\,
\Lambda_{Nm}(\vec{\rho} - \vec{\rho}_0)\,\psi_{p}(z),
\nonumber \\
\beta_{\vec{K}Q}\left(\vec{r}_e,\vec{r}_h\right) &=
\vec{K}\vec{R}_\perp + QZ +
\frac{e}{2\hbar}\left(\vec{B}\times\vec{\rho}\right)\vec{R}_\perp +
\frac{\delta}{2}\,\vec{K}'\vec{\rho} -
\frac{1}{2}\,\frac{\Delta_e-\Delta_h}{\Delta_e+\Delta_h}\,Qz,
\end{align}
where $Z=(z_e+z_h)/2$ (notice that for equal electron and hole masses $Z$
coincides with the exciton longitudinal center-of-mass coordinate).
With $L=N_0d$ and $S$ being the length and area, respectively, of the superlattice,
the exciton wave functions have the normalization property
$\langle \Psi_{\vec{K}QNmp} \mid \Psi_{\vec{K}'Q'N'm'p'} \rangle =
\delta_{\vec{K}\vec{K}'}\,\delta_{QQ'}\,\delta_{NN'}\,\delta_{mm'}\,\delta_{pp'}$.
In addition, we note that the longitudinal size of the exciton state
should be much smaller than the magnetic length, $\bar{z}<<a_B$,
in order to justify the employed above adiabatic separation
of the fast in-plane and slow longitudinal internal motions.

\section{Exciton states}\label{S:States}

To study the dependencies of the exciton absorption spectrum on the SL
parameters and the external fields we consider below the cases
amenable to analytical approximations for the binding
potential~(\ref{E:pot1}). To simplify the derivations, we adopt
$\vec{K}=K\vec{e}_x$ for the exciton in-plane momentum and
$\vec{F}=F\vec{e}_y$ for the external electric field,
which yields $\rho_0 = (\hbar/eB)\left[K-(M/{\hbar}B)F\right]$.
Below we distinguish two cases,
$\rho_0 \ll a_B \ll a_0$ and $a_B \ll \rho_0 \ll a_0$.

\subsection{The case $\rho_0 \ll a_B \ll a_0$}

Let us first derive an asymptotic expression for the potential
for $\rho_0 \simeq 0$, when the values of the in-plane exciton momentum
$K$ and electric field $F$ do not effect the relative longitudinal motion.
Since the optically active exciton transitions are proportional to
$|\Psi(\vec{r}_e = \vec{r}_h)|^2$~\cite{ell},
we have to set the magnetic quantum number $m=0$.
The potential $\bar{U}_N = \bar{U}_{N0}$ acquires the form of the triangular
quantum well,
\begin{equation}\label{E:pot1a}
\bar{U}_N (z;0) = \bar{U}_N (0;0) + eF_0|z|,
\end{equation}
where
\begin{align}\label{E:parameters1a}
& \bar{U}_N (0;0) = \frac{e^2\beta_N}{4\pi\varepsilon_0 \varepsilon a_B},
\;\;\;
  F_0 = \frac{e}{4\pi\varepsilon_0 \varepsilon a_B^2},
\nonumber \\
& \beta_N = - \sqrt{\frac{\pi}{2}}
\sum_{j=0}\frac{[(2j-1)!!]^2 [2(N-j)-1]!!}{2^{N+j}(N-j)!(j!)^2}.
\end{align}
With increasing $\rho_0$ the potential well becomes more shallow.
For $\rho_0 \ll a_B$ we obtain for the ground $N=0$ Landau states
\begin{equation}\label{tqw1}
\bar{U}_0(z;\rho_0) = \left( 1 - \frac{s}{2} \right) \bar{U}_0(0;0),
                    + (1-s) eF_0 |z|,
\end{equation}
where $s = \rho_0^2/(2a_B^2) \ll 1$. When $s$ approaches unity
($\rho_0 = \sqrt{2}a_B$), the potential becomes
\begin{equation}\label{tqw2}
\bar{U}_0(z;\rho_0) = A\bar{U}_0(0;0) + \exp(-1)eF_0 |z|,
\end{equation}
where
\begin{equation}\label{a}
A \simeq 2 \pi^{-3/2}
\int_0^{\infty} \frac{x{\rm e}^{-x^2}}{1+x}
\left(- \ln\left[(x-1)^2\right] + \ln x
      + 2 \ln\left[8 \left(1+\sqrt{2} \right)\right] \right) {\rm d}x \simeq 0.80.
\end{equation}

The solution for the ground $p=1$ state in a one-dimensional triangular
quantum well~(\ref{E:pot1a}) was originally obtained by
Kohn and Luttinger~\cite{kohnlutt}.
In general, a routine mathematics yields the orthonormalized optically
active even-parity wave functions $\psi_p(z)$ for
the ground and excited $p = 1,2,3,\ldots$ states in terms of the Airy
functions ${\rm Ai}(\xi)$~\cite{abram}. For the potential~(\ref{E:pot1a})
we obtain

\begin{align}\label{E:func1}
\psi_p(z) &= D_1(p){\rm Ai}\left(\frac{|z|-z_0}{z_1}\right), \;\;\;
D_1(p) = \left[ -2c_p z_1 {\rm Ai}^2(c_p) \right]^{-1/2},
\nonumber \\
z_0 & = \frac{W_{Np} -\bar{U}_N(0;0)}{eF_0}, \;\;\;
z_1 = \left[ \frac{\mu\,a_0 a_B^2}{2m_\parallel(Q)} \right]^{1/3},
\end{align}
where

\begin{equation}\label{E:ener1}
W_{Np} = \bar{U}_N (0;0)
- \frac{\hbar eB}{\mu}
  \left[ \frac{\mu\,a_B^2}{2m_{\parallel}(Q) a_0^2} \right]^{1/3}c_p
\end{equation}
are the corresponding eigenenergies, see Eq.~(\ref{E:relat1}).
The coefficients $c_p$ are the roots of the equation ${\rm Ai}'(c_p) = 0$ which yields
$c_1 = -1.02$, $c_2 = -3.25$, $c_3 = -4.82$, $\ldots$, and
${\rm Ai}(c_1) = 0.54$, ${\rm Ai}(c_2) = -0.42$, ${\rm Ai}(c_3) = 0.38$, $\ldots$.
For the potentials~(\ref{tqw1}) and (\ref{tqw2}), the analytical results for
the wave functions and the energies are transformed accordingly:
$\bar{U}_N(0;0)$ and $F_0$ are replaced by
$(1-0.5s)\bar{U}_0(0;0)$ and $(1-s)F_0$, respectively,
for the potential~(\ref{tqw1}), and by
$A\bar{U}_0(0;0)$ and $\exp(-1)F_0$, respectively,
for the potential~(\ref{tqw2}). The inter-level distances
$\Delta W_{Np}= W_{N,p+1} - W_{Np}$
become smaller in energy.

The parameter $z_1$ quantifying the extension of the exciton
wave function~(\ref{E:func1}) along the magnetic field
should respect the applied above continual and adiabatic
approximations. Therefore, the analytical
approximations~(\ref{E:func1}) and (\ref{E:ener1}) for the exciton quantum states
require fulfillment of the condition
\begin{equation}\label{E:cond1}
d \ll z_1 \ll a_B \ll a_0.
\end{equation}

\subsection{The case $a_B \ll \rho_0 \ll a_0$}

In this case the shift $\rho_0$ in the denominator of the integrand
in the potential~(\ref{E:pot1}) can be considered large compared to $\rho$
and $|z|$, and the potential can be approximated by

\begin{align}\label{E:oscpot}
\bar{U}_{Nm}(z;\rho_0) &=
\bar{U}_{Nm}(0;\rho_0) + \frac{1}{2}\,m_{\parallel}(Q) \Omega^2 z^2,
\nonumber \\
\bar{U}_{Nm}(0;\rho_0) &= - \frac{e^2}{4\pi\varepsilon_0\epsilon \rho_0},
\;\;\;
\Omega = \left[\frac{e^2}
{4\pi\varepsilon_0\epsilon m_{\parallel}(Q) \rho_0^3 }\right]^{1/2}.
\end{align}
The optically active even-parity orthonormalized wave functions
$\psi_p(z)$, $p=0,2,4,\ldots$ are the harmonic-oscillator functions
\begin{align}\label{E:herm}
\psi_p(z) &= D_2(p) \exp\left(-\xi^2/2\right) H_p(\xi),
\nonumber\\
D_2(p) &= \left( z_2\sqrt{\pi} p! 2^p \right)^{-1/2},
\;\;\;
\xi = \frac{z}{z_2},
\;\;\;
z_2 = \left( \frac{\hbar}{m_{\parallel}(Q) \Omega} \right)^{1/2},
\end{align}
where $H_p(\xi)$ are the Hermite polynomials~\cite{abram}. The corresponding
eigen energies are
\begin{equation}\label{E:oscen}
W_{Nmp} = \bar{U}_{Nm}(0;\rho_0) + \hbar \Omega \left(p+\frac{1}{2}\right).
\end{equation}
The analytical results~(\ref{E:herm}) and (\ref{E:oscen}) require the
fulfillment of the condition
\begin{equation}\label{E:cond2}
d \ll z_2 \ll a_B \ll \rho_0 \ll a_0,
\end{equation}
which justifies both the adiabatic and continual approximations employed.

\section{Spectrum of the exciton absorption. Results and discussion}\label{S:spectrum}

The exciton optical absorption coefficient can be calculated as follows~\cite{monzhil95}

\begin{equation}\label{E:coeff}
\alpha = \frac{n_0 \hbar\omega{\mathit \Pi}}{c\,\tilde{U}SNd},
\end{equation}
where $n_0$ is the refractive index, $c$ is the speed of light,
$\tilde{U}=\varepsilon_0 n_0^2 \mathscr{E}_{0}^2$ is the
energy density
of the optical radiation electric field
$\vec{\mathscr{E}}=\vec{\eta}\mathscr{E}_{0}\exp(-{\rm i}\omega t)$
with $\vec{\eta}$ being a unit polarization vector, and
\begin{equation}\label{E:rate}
{\mathit \Pi} = \frac{1}{t} \sum
\left| \frac{1}{{\rm i} \hbar} \int_0^t {\rm d}\tau \,
\langle \Psi | \mathscr{P}_{eh}(\tau) | \Psi_0 \rangle
\exp\left(\frac{{\rm i}}{\hbar}E\tau \right)\right|^2
\end{equation}
is the transition rate. The summation in Eq.~(\ref{E:rate}) is performed
over the exciton states with
the wave functions $\Psi = \Psi_{\vec{K}QNmp}(\vec{r}_e,\vec{r}_h)$
and the energies $E = E_{\vec{K}QNmp}$,
$\Psi_0 = \delta (\vec{r}_e - \vec{r}_h)$~\cite{ell}.
The matrix element $\langle \Psi | \mathscr{P}_{eh}(\tau) | \Psi_0 \rangle$
involving integration over $\vec{r}_e$ and $\vec{r}_h$ is
determined by the operator of electric dipole transitions~\cite{weiler}
\begin{equation}\label{dip}
\mathscr{P}_{eh}(t) =
\frac{2{\rm i}\hbar e \mathscr{E}_{0}}{m_0 E_g}\,
\left|\vec{\eta}\,\vec{p}_{eh}\right|\,
\delta_{\vec{q},\vec{P}}\,\cos(\omega t),
\end{equation}
where $E_g \simeq \hbar \omega$, $m_0$ is the mass of the free electron,
$\vec{p}_{eh}$ is a momentum averaged with the Bloch amplitudes
of the electron and hole bands, and $\vec{P}$ and $\vec{q}$
are the exciton and photon wave vectors, respectively.

Using the exciton wave functions derived in the previous section,
we obtain a spectral form of the absorption coefficient,
\begin{equation}\label{E:coeff1}
\alpha = \alpha^{(0)} \sum_{Nmp} f_{Nmp} \, \delta (E_{Nmp}-\hbar \omega),
\;\;\;
\alpha^{(0)} = \frac{\pi\hbar e^2}{\varepsilon_0 n_0 m_0 cSL},
\end{equation}
where
\begin{equation}\label{E:strength}
f_{Nmp} = \frac{2\left|\vec{\eta}\,\vec{p}_{eh}\right|^2}
               {m_0\hbar\omega a_0^3}\,G_{Nmp}
\;\;\; \mbox{and} \;\;\;
G_{Nmp}=a_0^3 \left|\Lambda_{Nm}(\rho_0)\psi_p(0)\right|^2
\end{equation}
are the oscillator strength density and the dimensionless oscillator strength,
respectively. The energies $E_{Nmp}$ are determined by Eq.~(\ref{E:energy}) for
conserved total momentum, $\vec{q}=\vec{P}$, as implied by the factor
$\delta_{\vec{q},\vec{P}}$ in Eq.~(\ref{dip}).

Analytical estimates, useful for the analysis of the exciton absorption,
can be obtained for the two distinct regimes considered in the
previous section.
For each regime we will discuss the dependencies on the electric $F$ and magnetic
$B$ field strengths for the binding energies
$E_{Nmp}^{(b)} = |W_{Nmp}|$, for the positions $E_{Nmp}$ of spectral lines
in the exciton absorption, and for the oscillator strengths $G_{Nmp}$.
We will also analyze the dependencies of these quantities on the angle $\vartheta$
which determines the direction of the photon wave-vector
$\vec{q} = K\vec{e}_x  + Q\vec{e}_z$, where $K=q\sin\vartheta$ and
$Q=q\cos\vartheta$, in the $x-z$ plane.

\subsection{The case $\rho_0 \ll a_B\ll a_0$}

In the case of small $\rho_0$ values
we may assume $K\simeq 0$, $F \simeq 0$ and estimate the oscillator strength as
\begin{equation}\label{E:max1}
G_{N0p} = \frac{1}{2\pi}\left( \frac{a_0}{a_B} \right)^2 \frac{1}{(-2c_p)}
          \left[ \frac{2m_\parallel(Q)}{\mu}
                 \left( \frac{a_0}{a_B} \right)^2 \right]^{1/3},
\;\;\;
p=1,2,3,\ldots,
\end{equation}
where the coefficients $c_p$ are the same as in Eq.~(\ref{E:ener1}).
The spectral peak positions $E_{N0p}$ can be
calculated from Eqs.~(\ref{E:energy}) and (\ref{E:ener1}) for $m=0$ and
$\vec{K} = \tilde{\vec{K}} = 0$.

\begin{figure*}[ht]
\includegraphics[width=0.99\textwidth]{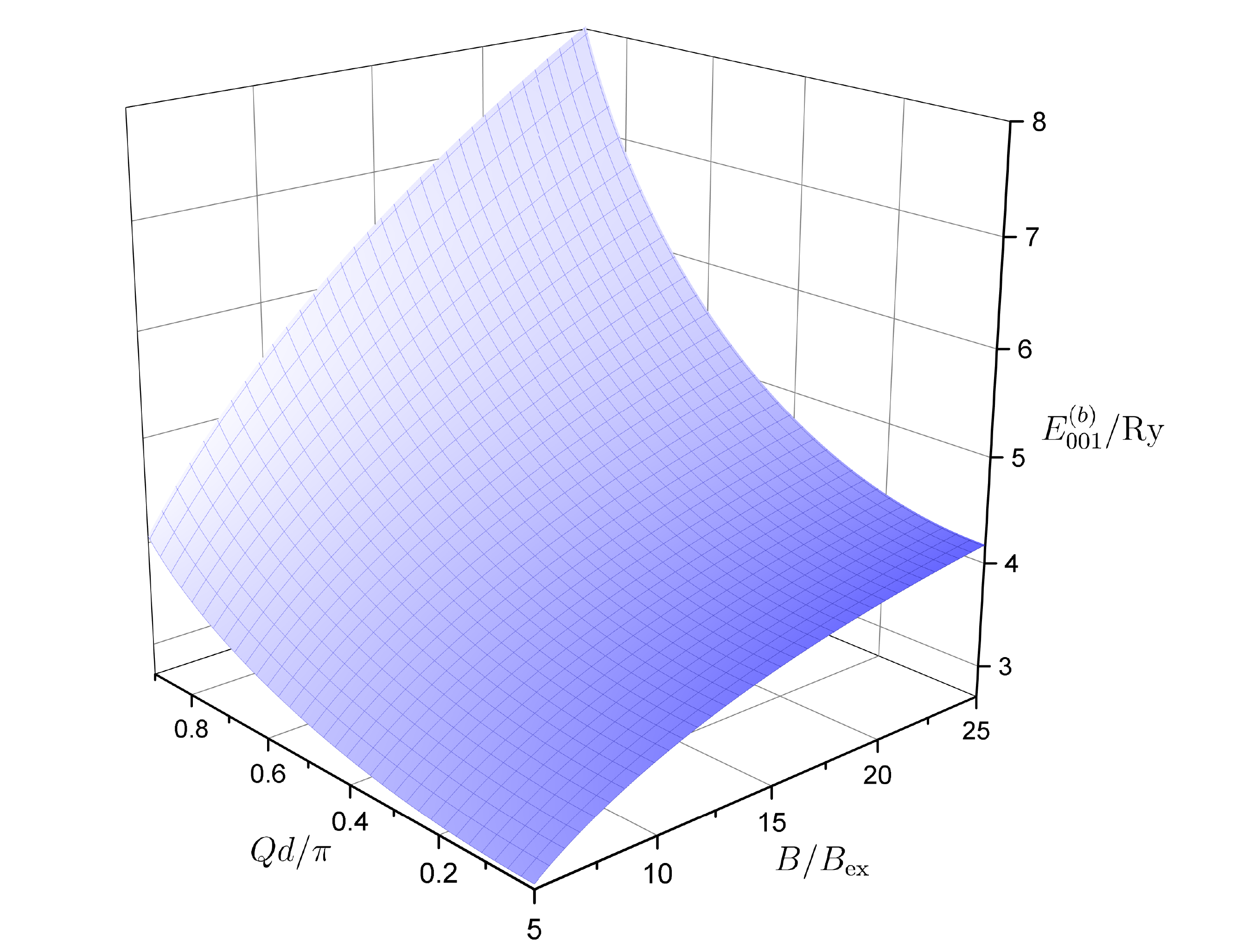}
\caption{The exciton ground state binding energy according
to Eq.~(\ref{E:ener1}) for $\mu/m_\parallel(0)=0.22$.
The energy (in units of the exciton Rydberg constant) is shown as a
function of the magnetic field strength (in units of
$B_{\rm ex} = \hbar/(ea_0^2)$) and the longitudinal component of the
exciton wave vector (in units of $\pi/d$).}
\label{fig1}
\end{figure*}

The account for small values of the in-plane exciton momentum and
the electric field has a minor effect on the exciton binding energies
and the absorption spectrum.
Notice that the first term in Eq.~(\ref{E:ener1})
significantly exceeds the second term, as it is reflected by the
condition $z_1 \ll a_B$ required for the adiabatic approximation.
It follows from Eq.~(\ref{E:ener1}) that, for a given Landau level
number $N$, both the binding energies
$E_{Np}^{(b)}=E_{N0p}^{(b)}$ and the energy gaps
$\Delta E_{Np}^{(b)}= E_{Np}^{(b)} - E_{N,p+1}^{(b)}$ between the
neighboring levels increase with increasing magnetic field
strength $B$ and decrease with increasing quantum number $p$.
It also follows that the energies $E_{Np}^{(b)}$ grow with the
value $Q$ of the exciton longitudinal momentum.
The binding properties are demonstrated in Fig.~\ref{fig1} for
the ground-state binding energy $E_{001}^{(b)}$
calculated as a function of $B$ and $Q$ according to Eq.~(\ref{E:ener1}).

With increasing $N$ the coefficient $\beta_N$ in Eq.~(\ref{E:pot1a})
decreases providing smaller values $|\bar{U}_N (0,0)|$ for the
depth of the triangular potential well. As a result, the
$N$-series of the exciton levels move to the region of smaller energies
for larger $N$.

\begin{figure*}[ht]
\includegraphics[width=0.99\textwidth]{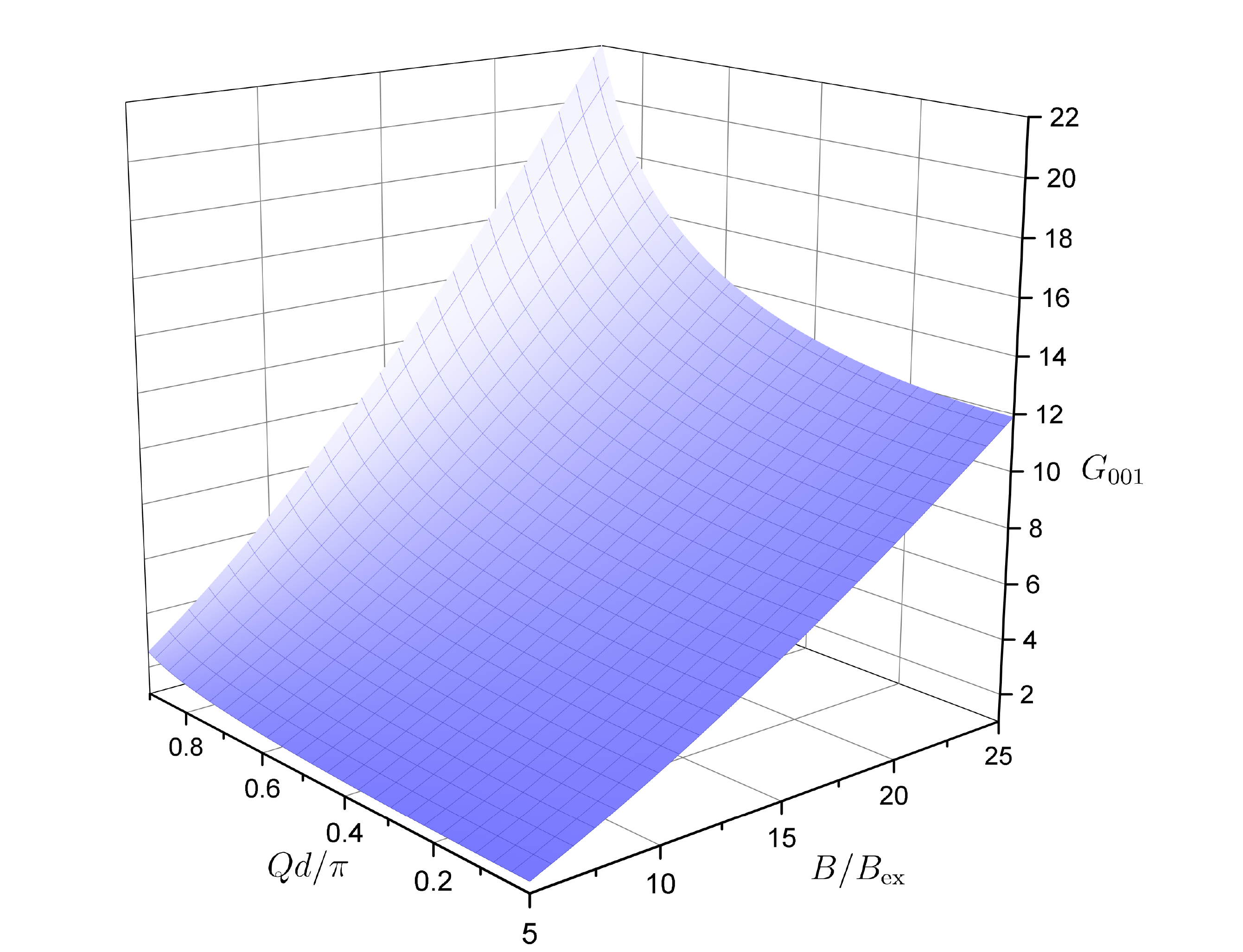}
\caption{The dependency of the dimensionless oscillator strength $G_{001}$ for
the ground peak in the exciton absorption on the
magnetic field strength $B$ and the longitudinal exciton wave vector $Q$.
Units for $B$ and $Q$ as in figure 1.}
\label{fig2}
\end{figure*}

Overall, the absorption spectrum~(\ref{E:coeff1}) comprises a sequence
of series of peaks corresponding to varying $p$ for different numbers $N$.
The peak positions in each series are adjacent from the lower energies
to the thresholds
\begin{equation}
\hbar \omega_N = \mathscr{E}_g + T(Q) + \frac{\hbar eB}{2\mu}\,(2N+1),
\label{thresholds}
\end{equation}
with the distance $\hbar eB/\mu$ between the neighboring thresholds
considerably exceeding the energy domain $\approx |\bar{U}_N(0,0)|$
of the series. With increasing magnetic field strength $B$ and the
longitudinal momentum $Q$ the $Np$ peaks shift toward higher
frequencies, while the distances between the neighboring
$Np$ and $N,p+1$ peaks increase with increasing magnetic field
strength and decrease with increasing longitudinal momentum.
For given $N$, the neighboring
peaks come closer to each other with increasing $p$, i.e. with
the peak positions approaching the thresholds~(\ref{thresholds}).
The peak intensities $f_{Np}$ increase as $\sim B^{4/3}$
with the magnetic field strength, and they decrease with growing
quantum number $p$. All the $N$-series of peaks are equal
in intensity.
Notice that, analogously to the binding energies,
the peak intensities increase as the longitudinal
momentum value $Q$ approaches the Brillouin zone boundary.
The above-discussed properties are illustrated in Fig.~\ref{fig2} by
the intensity $G_{001}$ of the ground-state exciton absorption
calculated according to Eq.~(\ref{E:max1}) as a function of
the magnetic field $B$ and the longitudinal momentum $Q$  for
$\mu/m_{\parallel}(0) = 0.22$.

\subsection{The case $a_B \ll \rho_0 \ll a_0$}

In contrast to the previous case, the quantity $\rho_0$ is now
not close to zero, and the in-plane total wave vector $K$ and
the electric field $F$ significantly influence the exciton binding
energies and the absorption spectrum.
The
oscillator strengths become

\begin{equation}\label{E:max2}
G_{Nmp} = \frac{1}{2\pi}\left( \frac{a_0}{a_B} \right)^2
          \frac{s^{2N+|m|}}{N!(N+|m|)!}\,\exp(-s)
          \left[\frac{m_{\parallel}(Q)}{\mu}
                \left(\frac{a_0}{\rho_0}\right)^3\right]^{1/4}
          \frac{\left[(p-1)!!\right]^2}{\sqrt{\pi}\,p!},
\end{equation}
where $s = \rho_0^2/(2a_B^2) \gg 1$, $N,|m| = 0,1,2\ldots$, $p=0,2,4,\ldots$.
The peak positions $E_{Nmp}$ are determined
by Eqs.~(\ref{E:energy}) and (\ref{E:oscen}).
To simplify the analysis of the effects of the electric field and exciton momentum,
we will consider the cases where $\rho_0$ is dominantly determined by either
$F$ or $K$ values.

\subsection*{B1. The case $a_B \ll \rho_0 \ll a_0$, $K \ll MF/(\hbar B)$}

\begin{figure*}[ht]
\includegraphics[width=0.99\textwidth]{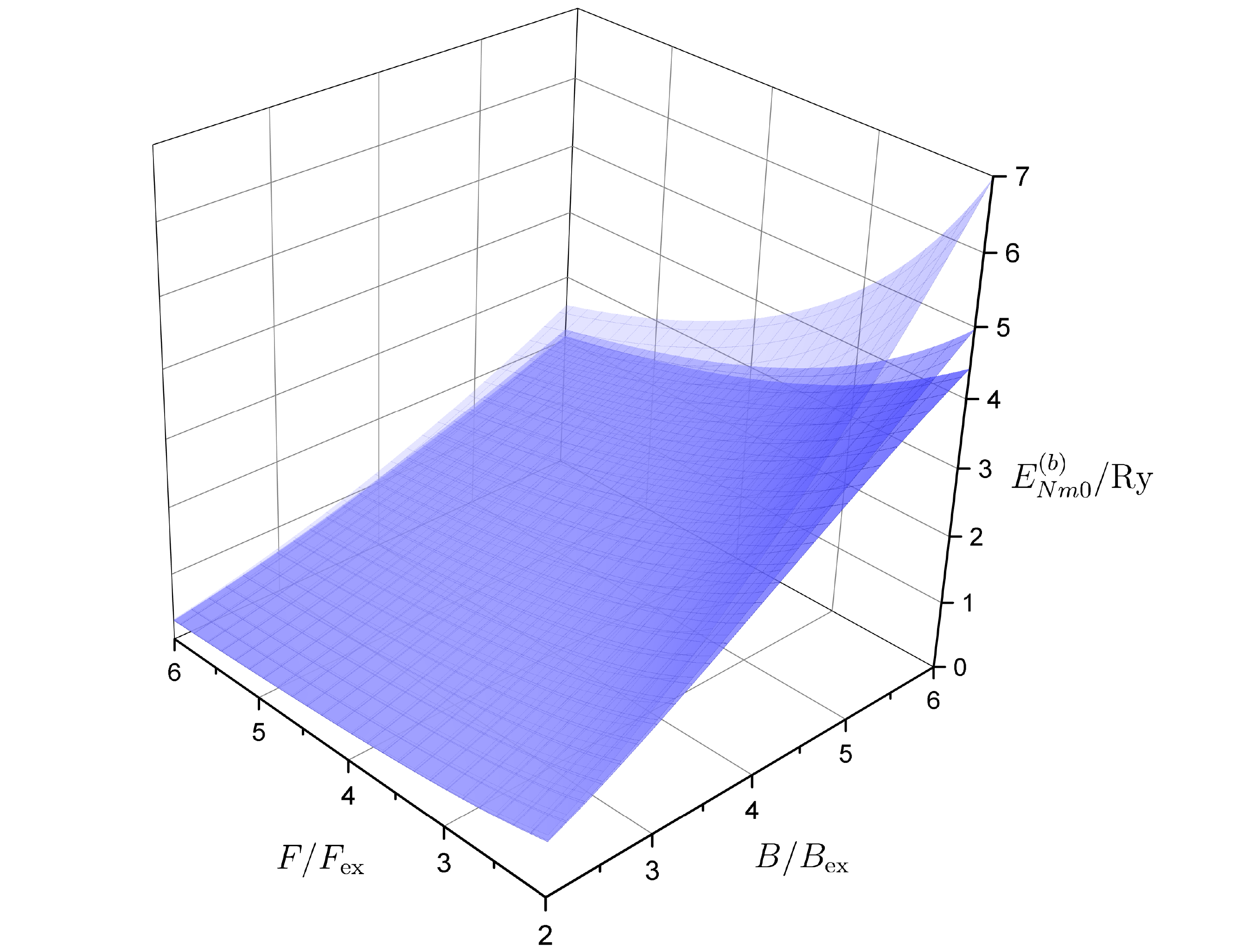}
\caption{The binding energies (in units of the exciton Rydberg constant)
of the exciton $p=0$ states. The energies are the same for all $N,m$ series of
states and calculated according to Eq.~(\ref{E:oscen}) for
$\mu/m_\parallel(0) = 0.22$ and $\mu/M=0.12$. The surface plots
show the energies as functions of the strengths of the external electric $F$
and magnetic $B$ fields (scaled with $F_{\rm ex} = {\rm Ry}/(ea_0)$
and $B_{\rm ex} = \hbar/(ea_0^2)$, respectively), for three different
values of the longitudinal wave vector: $Qd/\pi=0$ (bottom surface),
$Qd/\pi=0.45$ (middle surface), and $Qd/\pi=0.9$ (upper surface).}
\label{fig3}
\end{figure*}

\begin{figure*}[ht]
\includegraphics[width=0.99\textwidth]{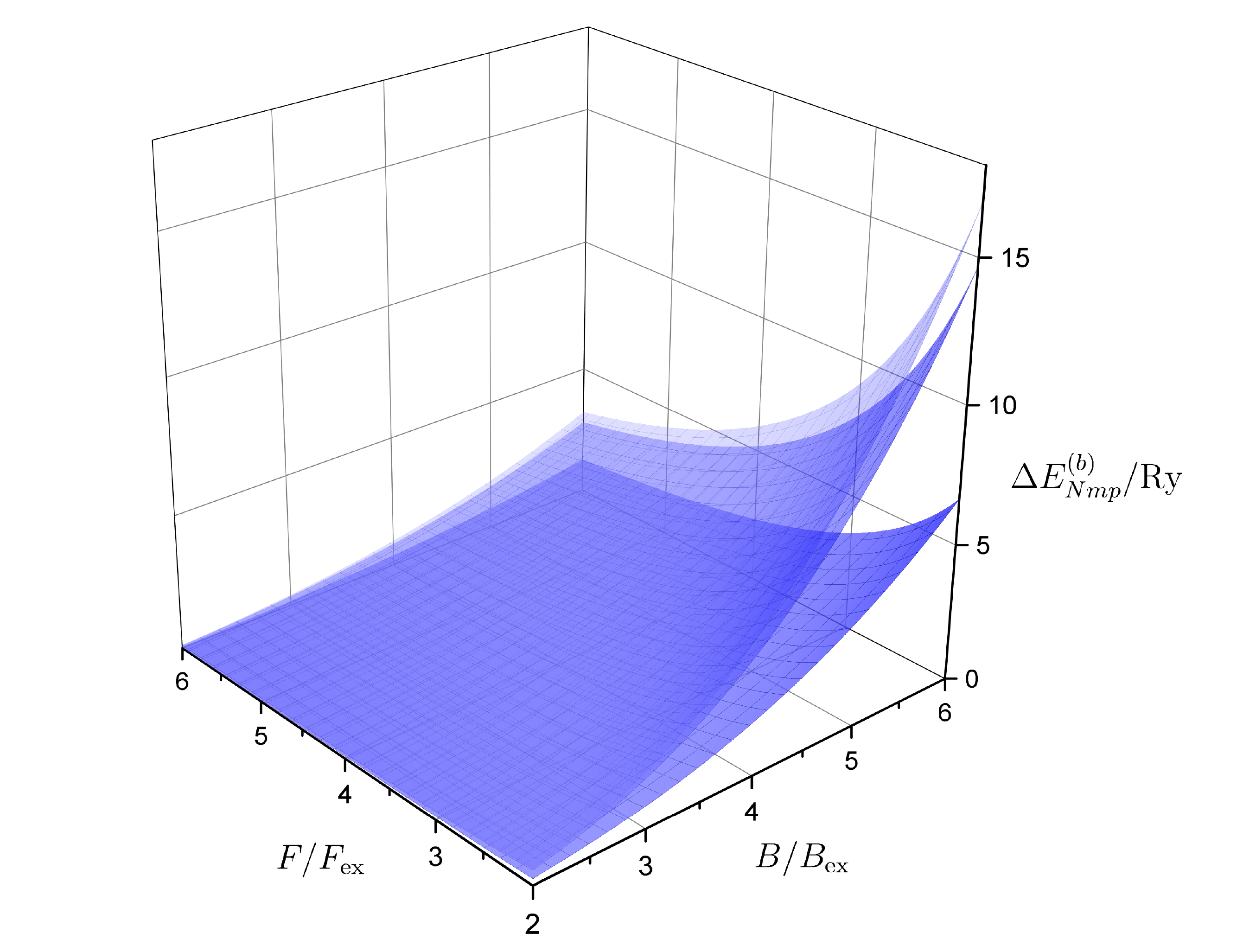}
\caption{Surface plots showing the distance between the neighboring exciton
absorption peaks (in units of the exciton Rydberg constant) as a function of
$F/F_{\rm ex}$ and $B/B_{\rm ex}$ for the same parameters as in figure ~\ref{fig3}.
The upper, middle and bottom surfaces correspond to the values $0$, $0.45$ and
$0.9$ of $Qd/\pi$, respectively.}
\label{fig4}
\end{figure*}

In this case the electric field influences the exciton absorption
to a much stronger extent than the in-plane momentum.
Because of the condition $z_2 \ll a_B$ required by the adiabatic
approximation, the binding potential~(\ref{E:oscpot}) and the
quantum energies~(\ref{E:oscen}) are dominated by the first terms
of the corresponding expressions. It follows then that, both the binding
energies $E_p^{(b)} = |W_{Nmp}^{(b)}|$ and the distances
$\Delta E^{(b)} = E_p^{(b)} - E_{p+2}^{(b)} = 2 \hbar \Omega$ between
the neighboring levels decrease with increasing the strength of the
electric field. Increasing the longitudinal momentum $Q$ increases the
binding energies $E_p^{(b)}$ and decreases the gaps
$\Delta E^{(b)}$. As in the previous case, the binding energies
and the gaps are larger for a stronger magnetic field. They scale
with the field strength as $E_p^{(b)}\sim B^2$ and
$\Delta  E^{(b)}\sim B^3$. Notice that in the regime considered
the binding energies do not depend on the quantum numbers $N$ and $m$.
In Fig.~\ref{fig3}, we show the ground-state ($p=0$) binding energies
(Eq.~(\ref{E:oscen})) as functions of the electric $F$
and magnetic $B$ field strengths for $\mu/m_\parallel(0) = 0.22$,
$\mu/M=0.12$. Fig.~\ref{fig4} shows the dependence on $F$ and $B$ for
the energy separation $\Delta E^{(b)}$
between the equidistant exciton levels in each series.

The lines in the exciton absorption spectrum~(\ref{E:coeff1})
arise at the frequencies corresponding to the different sets $Nmp$
of the quantum numbers and group for each pair $Nm$ in the series.
Notice that the line series $m\neq 0$ result from the presence of
an in-plane electric field. With increasing $p$ the series saturate,
coming from lower frequencies, towards the thresholds
\begin{equation}
\hbar\omega_{Nm} = \mathscr{E}_g +T(Q)
                  + \frac{\hbar eB}{2\mu} \Big( 2N + |m| + \delta\cdot m +1 \Big)
                  - \frac{MF^2}{2B^2},
\label{thresholds_b}
\end{equation}
which now depend on the field strengths.
The gaps $\sim \hbar eB/\mu$ between the thresholds significantly
exceed the energy ranges $|\bar{U}_{Nm}(0,\rho_0)|$ of the series
of absorption peaks. In each series the peak positions are equidistant
and separated by the interval $\Delta E_{Nmp}^{(b)} = 2\Omega$
which does not depend on $p$ in the harmonic approximation (\ref{E:oscen}).
Increasing the electric field strength shifts the $Nm$ series to lower
frequencies and narrows the series' energy ranges
$|\bar{U}_{N,m}(0,\rho_0)| \sim F^{-1}$.
The inter-peak separations in each series decrease with increasing
electric field and exciton longitudinal momentum,
$\Omega \sim \big( m_{\parallel}(Q)F^3 \big)^{-1/2}$.

\begin{figure*}[ht]
\includegraphics[width=0.99\textwidth]{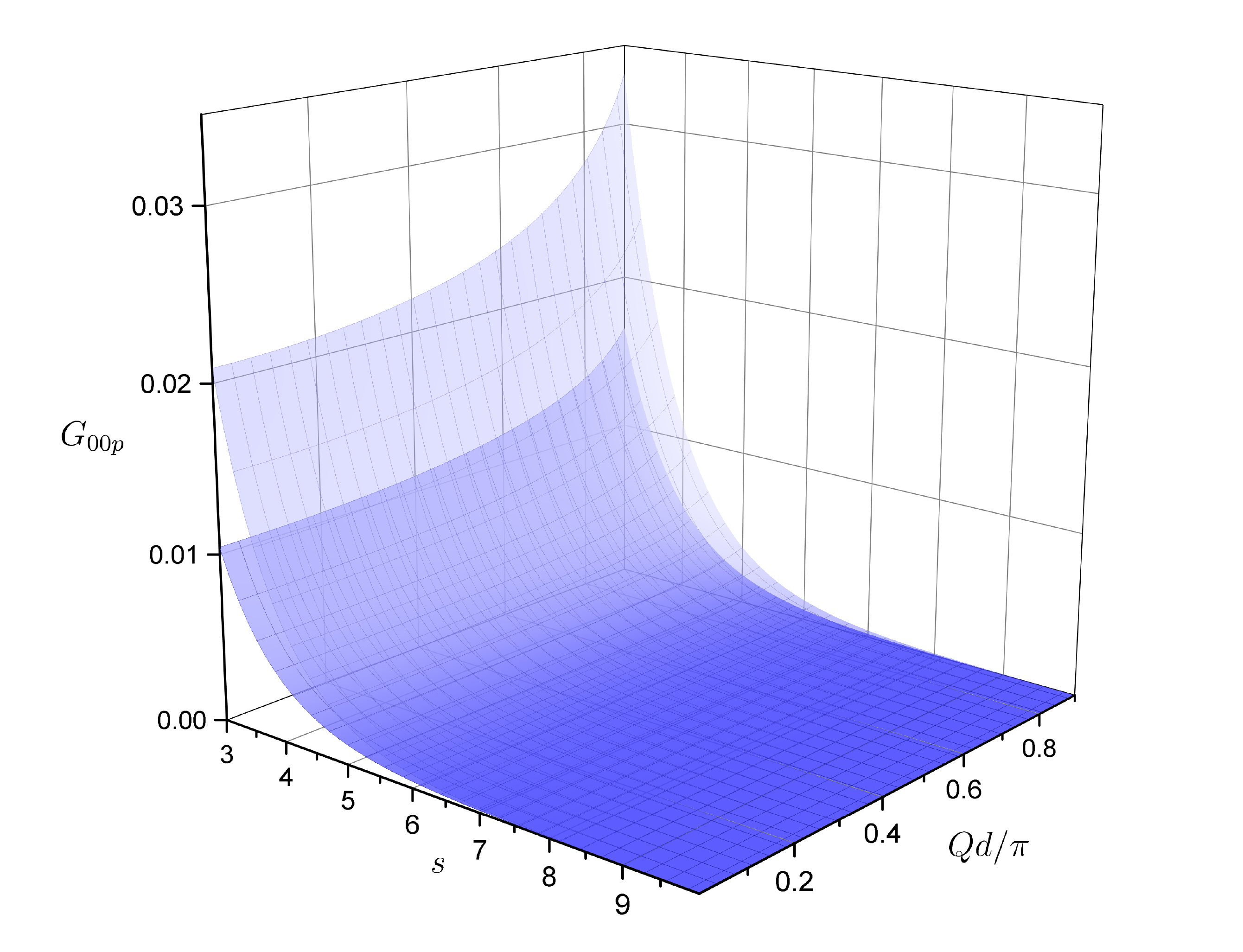}
\caption{The dimensionless oscillator strengths $G_{00p}$ of the transitions
to the ground $N=m=p=0$ (upper surface) and first excited
$N=m=0$, $p=2$ (bottom surface) exciton states.
The oscillator strengths are determined according to
Eq.~(\ref{E:max2}) for $\mu/m_\parallel(0)=0.22$ and
$B/B_{\rm ex} = 5$ and shown as functions of
the electric field strength
$F \sim s^{1/2}$ ($s = \rho_0^2/(2a_B^2)$) and the longitudinal
exciton wave vector scaled with the Brillouin zone boundary $\pi/d$.}
\label{fig5}
\end{figure*}

According to Eqs.~(\ref{E:strength}) and (\ref{E:max2}),
the oscillator strengths $f_{Nmp}$ increase with increasing magnetic
field strength $B$. Their dependence on the electric field strength $F$ is
non-monotonic for all except the ground ($N=m=0$) series.
For weak electric fields, $s \ll 1$, the ground series which is optically
active at $F=0$ reduces in intensity as $F$ increases, whereas the
series with $m\neq 0$ become more intense.
As $F$ increases in the range of moderate, $s\simeq 1$,
and large, $s\gg 1$, strengths, the oscillator strengths in all series decrease.
The maximal intensities in optical absorption increase with
the exciton wave number $Q$ approaching the Brillouin zone boundary.
The oscillator strengths decrease with increasing $p$.
For illustration, we show in Fig.~\ref{fig5} the oscillator strengths
$G_{00p}$ as functions of $F$ and $Q$ for the exciton absorption
to the ground $p=0$ and first excited $p=2$ states. They are determined
according to Eq.~(\ref{E:max2}) with the choice $\mu/m_\parallel(0) = 0.22$
and $(a_0/a_B)^2=5$.

\subsection*{B2. The case $a_B \ll \rho_0 \ll a_0$, $K \gg MF/(\hbar B)$}

In this situation the in-plane exciton momentum dominates the electric field
effects
influences of the exciton absorption. For better transparency of the effect
of in-plane exciton motion we put the length of the photon wave vector
to be close to the Brillouin zone boundary, and the electron and hole minibands
to coincide with each other. The exciton absorption coefficient $\alpha$,
peak positions $E_{Nmp}$ and the oscillator strength $f_{Nmp}$ can then
be derived from Eqs.~(\ref{E:coeff1}), (\ref{E:strength}), (\ref{E:oscen}),
(\ref{E:max2}) with account for the relations
\begin{align}
K &= q\sin\vartheta, \,\;\;
Q  = q\cos\vartheta, \;\;\;
q  = \pi/d, \;\;\;
\Delta_e = \Delta_h \equiv \Delta_0,
\nonumber \\
\rho_0 &= \frac{\pi a_B^2}{d}\,\sin\vartheta, \;\;\;
\Omega  = \frac{\hbar}{a_B^2 \mu}\,\frac{d}{a_B}
\left[
\frac{1}{\pi^3 }\,\frac{\mu}{m_{\parallel}(0)}\,\frac{d}{a_0}\,
\frac{\cos \left( \frac{\pi}{2}\cos\vartheta \right)}{\sin^3 \vartheta}
\right]^{1/2},
\label{relations}
\end{align}
where $\vartheta$ is the angle between the photon wave vector $\vec{q}$
and the SL axis.

\begin{figure*}[ht]
\includegraphics[width=0.99\textwidth]{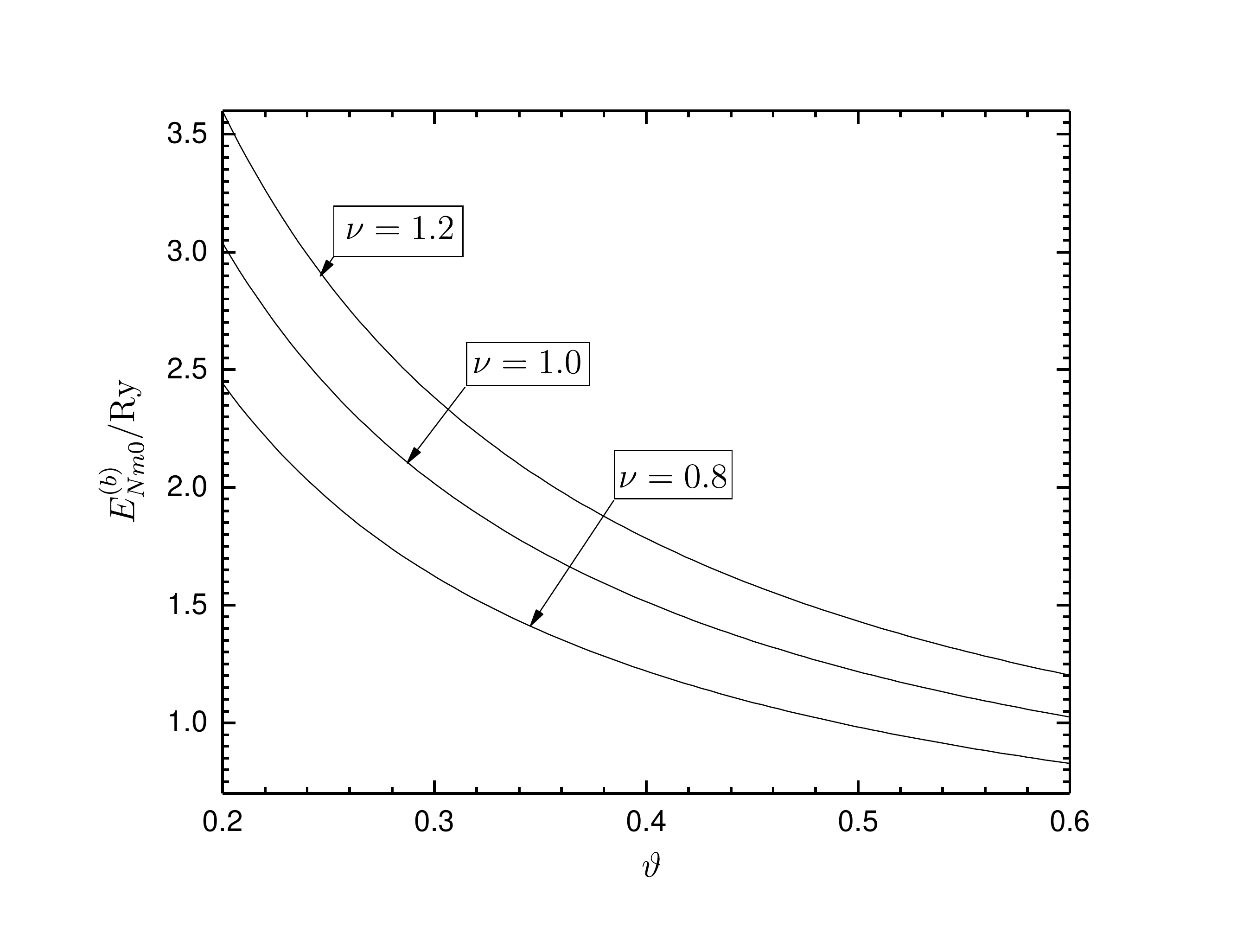}
\caption{The dependence of the exciton binding energies
$E_{Nm0}^{(b)}$ (in units of the exciton Rydberg constant) on the
angle $\theta$ between the photon wave vector and the SL axis.
The binding energies are the same for all $N$ and $m$ and determined
according to Eq.~(\ref{E:oscen}). The results correspond to
$\mu/m_\parallel(0)=0.22$ and three different values for $\nu = a_0 d/a_B^2$
(indicated near the curves).}
\label{fig6}
\end{figure*}

\begin{figure*}[ht]
\includegraphics[width=0.99\textwidth]{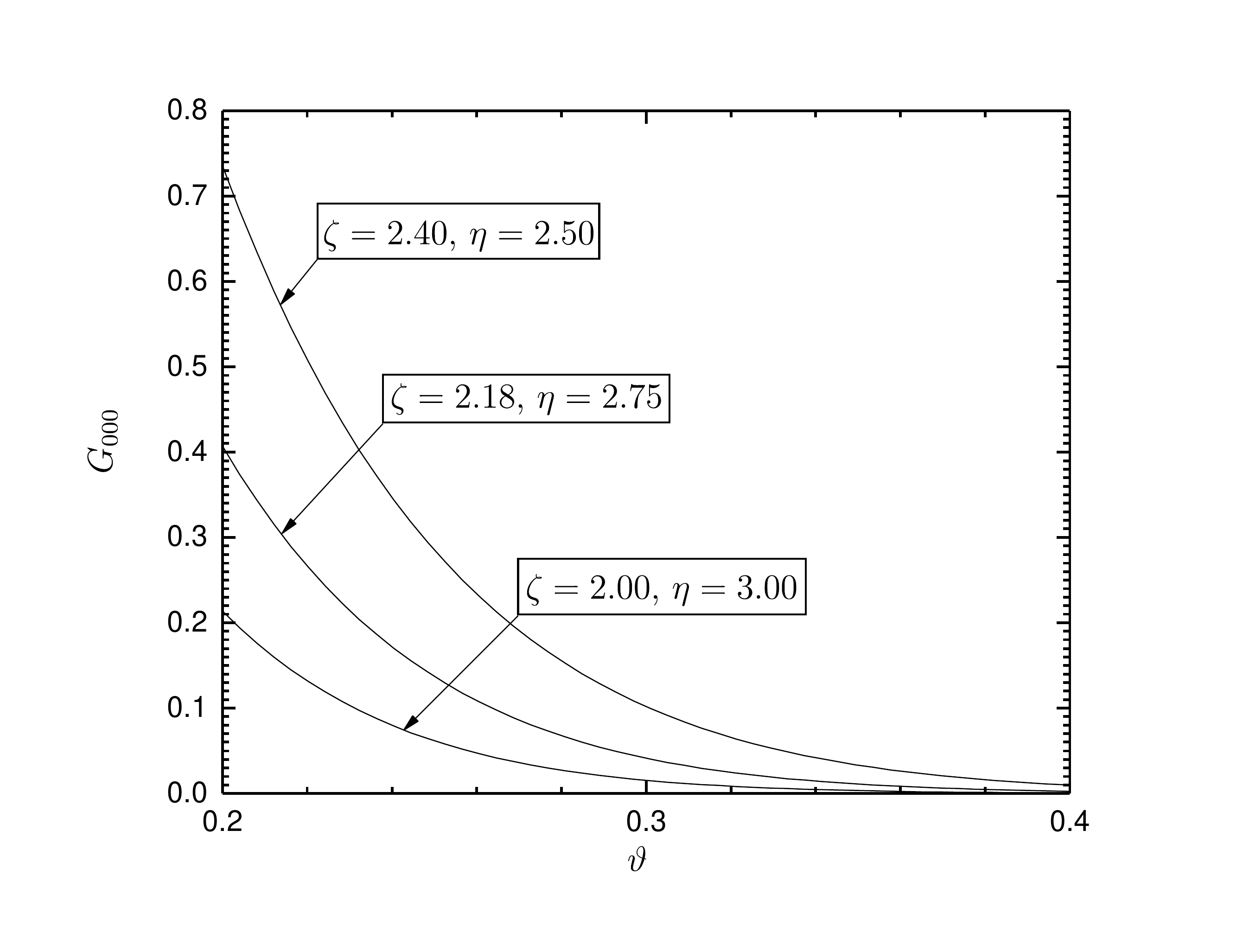}
\caption{The dimensionless oscillator strength $G_{000}$ for the ground peak
of the exciton absorption as a function of the angle $\vartheta$ between the
photon wave vector and the SL axis. The oscillator strength is determined
according to Eq.~(\ref{E:max2}) for $\mu/m_\parallel(0)=0.22$ and different
pairs of the parameters $\zeta$ and $\eta$ as indicated.}
\label{fig7}
\end{figure*}

The analysis shows that deviation of the vector $\vec{q}$ from the SL axis
results in decreasing binding energies
$E_{Nmp}^{(b)} = |W_{Nmp}|$ (cf. Eq.~(\ref{E:oscen})) and the
energy gaps $2\hbar\Omega$ between the neighboring $p$-states.
The peak positions $E_{Nmp} = E_{Nm} - E_{Nmp}^{(b)}$ shift towards
higher energies, though the thresholds
\begin{equation}
E_{Nm} = \mathscr{E}_g
       + 2\Delta_0 \sin^2 \left(\frac{\pi}{4}\cos\vartheta \right)
       + \frac{\hbar eB}{2\mu}(2N + \mid m \mid +\delta m +1),
\label{thresholds_c}
\end{equation}
become lower in energy.
The oscillator strengths $f_{Nmp}$ (see Eqs.~(\ref{E:strength}) and (\ref{E:max2}))
reduce in magnitude.
The numerical results are presented in Figs.~\ref{fig6} and \ref{fig7}.
Fig.~\ref{fig6} shows the ground state binding energy $E_{Nm0}^{(b)}$ calculated
according to Eq.~(\ref{E:oscen}) for $\mu/m_\parallel(0)=0.22$ and different
values $\nu = a_0 d/a_B^2$. Fig.~\ref{fig7} shows the oscillator strength $G_{000}$
of the transition to the ground exciton state $N=m=p=0$ calculated
according to Eq.~(\ref{E:max2}) for the same $\mu/m_\parallel(0)$ ratio
and different pairs of the values $\zeta = a_0/a_B$ and $\eta = a_B/d$.
A significant effect of the external electric and magnetic fields and of the
SL parameters studied in the figures is related to the condition~(\ref{E:cond2})
in which
\begin{equation}
z_2 = a_B
\left[\pi^3\,\frac{\mu}{m_{\parallel}(0)}\,\zeta \eta^3
\cos \left( \frac{\pi}{2}\cos\vartheta \right)
\sin^3\vartheta \right]^{1/4}.
\label{z_2}
\end{equation}

In the presence of crossed electric and magnetic fields, the exciton in SL
displays an {\em inversion effect}: changing orientation of any of the three vectors,
$\vec{F}$, $\vec{B}$ and $\vec{K}=\vec{q}_\perp$, to the opposite ones results in
a change of the exciton states and absorption spectrum. Originally this phenomenon
has been studied in connection with the Stark effect in the presence of magnetic
fields in the CdS bulk material~\cite{thomhopf}. According to the analytical results
presented here, the dependence of the peak positions $\hbar\omega = E_{Nmp}$ on
the orientation of the vectors $\vec{F}$, $\vec{B}$ and $\vec{K}$ follows from
Eqs.~(\ref{E:energy}) and (\ref{E:oscen}).
A particularly important prediction of our analysis is the existence of a critical
electric field $F_{\rm cr}=\hbar BK/M$ for which
$\rho_0 = 0$ and the distances between the neighboring $p$ and $p+1$ spectral
peaks reach a maximum determined by Eq.~(\ref{E:ener1}). This implies that
by varying the strength of the electric field $F$ in an experiment and
measuring the maximum distance $\Delta\omega$ between the ground $p=1$ and first
excited $p=2$ exciton peaks one can determine the effective mass $m_\parallel(Q)$
satisfying the equation
\begin{equation}
\Delta\omega = \frac{eB}{\mu}
\left( \frac{\mu}{2 m_{\parallel}(Q)} \frac{a_B^2}{a_0^2} \right)
(c_1 - c_2).
\label{eff_mass-exp}
\end{equation}
In this way it is possible to trace experimentally the dependence
on the longitudinal exciton momentum $Q$ and establish
the dispersion law for the exciton in the SL.
The latter parameter which is difficult to determine theoretically affects strongly
the electronic, optical and transport SL properties.

As originally pointed out in Ref.~\cite{suris},
the applicability of the analytical approach employed here
is restricted to the conditions~(\ref{E:cond1}) and (\ref{E:cond2}) which
involve the SL parameters $\Delta_{e,h}$, $d$, $\mu$ and the magnetic field
strength $B$. In order to meet these restrictions for SL which can be
experimentally probed, we consider the GaAs/Al$_x$Ga$_{1-x}$As SL
for $x=0.3$ with period $d=2$~nm in a magnetic field $B=20$~T. This SL has
the following parameters~\cite{harr}: $\mathscr{E}_g = 1.52$~eV,
$\mu = 0.06m_0$, $M = 0.5m_0$,
and ${\rm Ry} = 4.7$~meV. Notice that for the chosen SL the
conditions~(\ref{E:cond1}) and (\ref{E:cond2}) imply the longitudinal
effective mass to significantly exceed
the reduced mass, $m_\parallel(Q) \gg \mu$, as well as the minibands
$\Delta_{e,h}$ to be narrow and the period $d$ to be short. On the other hand,
the minibands become wider with decreasing $\mu$ and $d$,
$\Delta_{e,h} \sim \mu^{-1}d^{-2}$.
This conflict can be softened by fabrication of a SL with an optimal
relation between the barrier and well widths~\cite{suris}. For our analytical
estimates we adopt the ratio $\mu/m_\parallel(0)=0.22$ as the SL parameter,
and consider three regimes
corresponding to the cases A, B1 and B2, respectively.

\subsection*{1. $\rho_0 \ll a_B$}

We may set $F = K = Q = 0$ neglecting thereby the effects of the electric
field and exciton momentum on the exciton states and absorption.
Then the value of the ground state binding energy estimated according to
Eq.~(\ref{E:ener1}) is $E_{001}^{(b)} = |W_{01}| = 11.9$~meV.
Notice that in the absence of a magnetic field bound exciton states
do not exist~\cite{kohnlutt}.
The corresponding dimensionless oscillator strength given by
Eq.~(\ref{E:max1}) is $G_{001}=1.0$, and the energy separation between the neighboring
Landau series is $\Delta_L = \hbar eB/\mu = 38.3$~meV.

\subsection*{2. $\rho_0 \gg a_B$, $K=Q=0$, $F \neq 0$}

As in the previous case, we assume the exciton momentum to be zero. However, the
electric field now does not vanish, and we pick a value $F = 9.0$~kV/cm satisfying
the conditions above. The ground exciton state binding energy determined
according to Eq.~(\ref{E:oscen}) is $E_{Nm0}^{(b)} = |W_{Nm0}| = 11.3$~meV.
Eq.~(\ref{E:energy}) allows to estimate a threshold red shift due to the
applied electric field, $\Delta_F = MF^2/(2B^2) = 2.9$~meV, whereas
Eq.~(\ref{E:max2}) yields $G_{000} = 0.25$ for the dimensionless oscillator
strength of the transition to the ground exciton state.

\subsection*{3. $\rho_0 \gg a_B$, $F = 0$, $K \neq 0$}

In order to address the interplay between the longitudinal
$Q = q\cos\vartheta$ and the transverse $K = q\sin\vartheta$ components of the
exciton wave vector we set $F = 0$ and $q = \pi/d$. The results acquire a most
simple form for the equal electron and hole miniband widths
$\Delta_e =\Delta_h$. As already pointed out, the conditions~(\ref{E:cond2})
hold for a narrow domain of the angles $\vartheta$. Nevertheless, variations
of the ground state binding energies and oscillator strengths with
$\vartheta$ can be well distinguished in an experiment. The estimates
according to Eqs.~(\ref{E:oscen})and (\ref{E:max2}) yield
$E_{Nm0}^{(b)} = 10.01~meV$, $G_{000} = 0.37$ and
$E_{Nm0}^{(b)} = 13.42$~meV, $G_{000} = 1.05$ for the angles
$\vartheta = 0.20$ and $\vartheta = 0.15$, respectively.

We believe that our analytical results are in line with the current
technological progress in fabricating SLs with a narrow
miniband and a short period. The corresponding numerical estimates
demonstrate that the fine structure of the exciton
magnetoelectroabsorption can be experimentally detected.

\section{Conclusions}\label{S:concl}

We have studied the quantum states and optical absorption for
an exciton in a semiconductor SL subject to external
crossed electric and magnetic fields directed perpendicular and
parallel to the SL axis, respectively. Analytical results
have been obtained for the exciton wave functions and energies
as well as for the transition energies and oscillator strengths
of the absorption spectra. We have considered the case where the
magnetic length considerably exceeds the SL period but is much
smaller than the exciton Bohr radius. Here, an adiabatic separation
is possible for the two types of internal exciton motions: the fast
in-plane motion and the slow longitudinal motion governed by the
external magnetic and electric and the internal Coulomb fields, respectively.

We have focused on the fine structure of the exciton energies
related to the bound electron-hole states in effective one-dimensional
potentials which are triangular-type and oscillator-type for
a weak and a strong effective electric field, respectively.
The latter field is a combination of the external electric field and
a field induced by the exciton center-of-mass motion in the external
magnetic field. It was shown that an increase of the magnetic field
leads to an increase of the binding energies, of the
energy separations between the states, and of the oscillator strengths
of optical transitions. In the opposite positive case, with increasing effective electric
field the latter energy and spectral quantities decrease.
An inversion effect for the exciton
absorption with changing external electric and magnetic fields and
switching the directions of the in-plane exciton momentum, has been
highlighted. A novel interplay of external crossed
fields, transverse and especially longitudinal
centre-of-mass and internal exciton motions has been found to occur.
The analytical results and numerical estimates
obtained can be useful for extending our knowledge on the excitonic
magneto-electro-states relevant to micro- and opto-electronics.

\end{document}